# Amorphous and ordered states of concentrated hard spheres under oscillatory shear

Nick Koumakis[*,1,2], John F. Brady[3], George Petekidis[1]

[1]FORTH/IESL & Dept. of Mat. Sci. & Tech., Univ. of Crete, 71110, Heraklion, Greece
[2]SUPA, School of Physics and Astronomy, The University of Edinburgh, Edinburgh, EH9 3JZ, UK
[3]Div. of Chem. and Chem. Eng., California Institute of Technology, Pasadena, CA 91125, USA
*Corresponding Author: nick.koumakis@ed.ac.uk

## Abstract

Hard sphere colloidal particles are a basic model system for general research into phase behavior, ordering and out-equilibrium glass transitions. Experimentally it has been shown that oscillatory shearing of a monodisperse hard sphere glass, produces two different crystal orientations; a Face Centered Cubic (FCC) crystal with the close packed direction parallel to shear at high strains and an FCC crystal with the close packed direction perpendicular to shear at low strains. Here, using Brownian dynamics simulations of hard sphere particles, we have examined high volume fraction shear-induced crystals under oscillatory shear as well as the same volume fraction glass counterparts. We find that, while the displacements under shear of the glass are isotropic, the sheared FCC crystal structures oriented parallel to shear, are anisotropic due to the cooperative motion of velocity-vorticity layers of particles sliding over each other. These sliding layers generally result in lower stresses and less overall particle displacements. Additionally, from the two crystal types, the perpendicular crystal exhibits less stresses and displacements at smaller strains, however at larger strains, the sliding layers of the parallel crystal are found to be more efficient in minimizing stresses and displacements, while the perpendicular crystal becomes unstable. The findings of this work suggest that the process of shear-induced ordering for a colloidal glass is facilitated by large out of cage displacements, which allow the system to explore the energy landscape and find the minima in energy, stresses and displacements by configuring particles into a crystal oriented parallel to shear.

## Introduction

Suspensions of colloidal hard spheres have been used as model systems [1], to offer insights into many fundamental phenomena of condensed matter physics such as equilibrium phase transitions [2], but also out of equilibrium states such as glasses, gels and attractive glass transitions [3, 4]. The defining feature of a colloidal glass is low particle mobility within a disordered solid, orders of magnitude smaller than that of a corresponding fluid. Colloids provide a valuable model system for the examination of the glass transition due to the time and length scale accessibility [5]. Thus, considerable insight is gained into features such as motion cooperativity and dynamic heterogeneity [6, 7]. However, because of the large particle size, colloids form soft solids which flow under the application of shear [8], where non linear rheological properties can be studied.

For hard spheres, there is a single parameter which completely determines the phase behavior at rest, the particle volume fraction φ. Hard spheres exhibit a liquid phase at volume fractions beneath 0.494, where the particles are free to diffuse and explore the whole volume available. A liquid-crystal coexistence phase is observed at volume fractions ranging from 0.494 to 0.545 and a fully crystalline structure from 0.545 to about 0.58 [9, 10]. The increase of entropy and free volume are the driving forces behind the crystallization of hard sphere colloids. The ordered crystal structure allows each particle a larger volume of motion around their lattice site in comparison to a disordered state of the same volume fraction. Even though the particles are in an ordered state, the increase of entropy due to individual motion and free volume is greater than the reduction of the total configurational entropy [9]. The crystal structures initially assembled to when left at rest are a mixture of face-centred cubic (FCC) and hexagonally close packed (HCP) regions that are randomly oriented [11]. If particles increase in polydispersity, crystallization dynamics become slower as different particles sizes are not easily accommodated in the crystal lattice. At about 10% polydispersity, the distribution of sizes is too large and crystallization is suppressed [9, 12, 13].

Further increase of the volume fraction, for φ above 0.58, the particles are unable to move into the entropically favorable crystal positions due to space restrictions and become trapped in a dynamically arrested glass phase [4, 14, 15]; Molecular dynamics simulations have brought up a discussion on the definition of the glass transition volume fraction and the appearance of crystallization in the glass regime [16]. While highly polydisperse hard sphere systems are unable to crystallize, the arrested glassy state at high φ remains unchanged. Above the glass transition, although long-range diffusion is essentially frozen [17, 18], dynamic heterogeneities may allow slow non-diffusive relaxations related with activated hoping mechanisms [6, 19]. By observing the positions of the fastest and slowest particles [7] it was found that just below the glass transition the motions of the fast-moving particles were strongly correlated spatially in clusters. As the glass transition was approached these domains grew in size, although when entering the glass phase, the average size of these clusters was reduced, providing a dynamic signature of the glass transition [20, 21]. The dynamic heterogeneities near and above the glass transition have also been discussed [22] and shown to have large distance spatial correlations of a few particle diameters which increase with φ, although showing no divergent behavior near or above the glass transition.

Rheologically, hard sphere glasses are an interesting system, generally exhibiting solid like response at rest, while showing complex features under non-linear shear. Earlier work [23] studied the linear elasticity of hard sphere dispersions near and above the glass transition, concerning the amorphous and shear induced ordered states, finding a power law increase of elasticity as a function of the distance from maximum packing, while additionally successfully applying the Mode Coupling Theory for the frequency dependence of the moduli. Creep and recovery experiments [24] as well as some oscillatory shear measurements coupled with DLS echo [8, 25] were used to examine the nonlinear properties of a polydisperse hard sphere glass and concluded that it can tolerate a high amount of strain (10%-15%) before yielding irreversibly as related to breaking of the hard sphere entropic cages. Moreover, a recent study presented an overview of the various rheological properties of concentrated hard sphere suspensions in comparison to their softer counterparts [26], covering linear viscoelasticity and non-linear steady and oscillatory measurements, as well as providing an experimental rheological signature of the glass transition.

There has also been recent scientific activity in the experimental determination of the microscopic properties of concentrated hard sphere glasses under shear. With the use of confocal microscopy coupled with steady shear, a sub linear power law dependence for the shear induced diffusion coefficient with increasing shear rate[27]. Moreover, recent work [28] relates the shear banding instability in high concentration glasses to shear concentration coupling, while mapping the occurrence of banding with applied rate and volume fraction. Through microscopy and steady shear localized irreversible shear transformation zones were identified and their formation energy and topology were determined[29], while other work [30] related the microscopic motions of the particles under shear to a modified Stokes-Einstein relation which replaces the thermal energy with shear energy. Additionally, the dynamic heterogeneities under steady shear have been analyzed for a super cooled liquid [31], while also recently studied in terms of creep flow [32], showing an increase of the heterogeneous regions as a function of strain. The source of the transient overshoot for steady rate start-up has been associated with structural changes of the glassy cage [33-35], while a secondary yield strain was found in simple hard sphere glasses [36], explained through distinct yielding mechanisms at different time scales. In a collaborative effort on glasses under shear, the relaxation of stresses after the application of flow was examined with theory, simulations and various experiments [37], showing that finite residual stresses are governed by the preshear rate through long-lived memory effects.

In a monodisperse system of hard spheres, the addition of shear has been found to induce ordering. While crystallization in quiescent hard sphere systems occurs within hours or days as randomly oriented crystallites, under oscillatory shear, crystallization may occur within a few minutes and is oriented with the flow[38]. Depending on the studied volume fraction, the created crystal under shear may be transient ($\varphi<0.454$), partial ($0.454<\varphi<0.545$) or monocrystalline ($\varphi>0.545$) [39], while it flows in complicated fashion with crystal layers sliding one over the other [40]. It has been found for dense suspensions of hard spheres ($\varphi>0.545$) under oscillatory strain that two different orientations of crystals arise depending on the strain of oscillation; while high strains ($\gamma_0>50\%$) produce a monocrystal with the closed packed direction parallel to shear, while low strains ($\gamma_0<50\%$) produce a distribution of crystallites with orientations around a crystal with the closed packed direction perpendicular to shear [38, 39, 41].

At low strains, the crystal has been found to have some sort of polycrystallinity as the sample would not be fully oriented in the perpendicular direction, but would have grains which had a preferred direction which was perpendicular to the shear direction[39]. At high strains the sample would reorient parallel to shear and the polycrystallinity would disappear. In the case of the cone and plate geometries, crystallization manifested only as a crystal with the closed packed direction parallel to shear, possibly due to spatial constrictions for the perpendicular crystal in a rotational configuration [23]. In [42] there is a discussion on the physical interpretation of the change of orientation from perpendicular at low strains to parallel to shear at high strains.

Static light scattering in addition to oscillatory shear was used to examine the time dependent growth of a crystal under shear, while exploring the parameters for optimum crystallization [43]. Light scattering echo and optical microscopy were used to examine hard sphere glasses under oscillatory shear strain [41], finding the crystal growth time scales for high φ glass. Two dimensional simulations of repulsive and attractive systems under shear revealed an optimal shear rate where crystallization speed is enhanced over the one at rest [44]. Brownian dynamics simulations were carried out on crystal nucleation with the application of steady shear and an umbrella sampling technique, which found that shear suppresses nucleation and leads to a larger critical nucleus, which promotes crystallization [45, 46]. Other work[47] has examined the shear induced ordering in an attractive gel with microscopy and light scattering echo and proposed a model for particle escape time and crystallization. Through simulations and confocal microscopy, investigations of the real-space structure revealed four distinct oscillatory shear-induced phases in hard-sphere fluids [48]. Furthermore, a range of different simulation techniques to examine the mechanism of shear induced ordering on jammed systems and deduced that ordering occurs as shear pushes the system to lower energy minima[49]. There is an interesting general review on colloidal shear induced ordering[50], while a more recent review of nucleation in 2D and 3D crystals at rest and under external fields [51] has also been published.

Previous work on the rheological behavior of shear induced crystallization in hard sphere glasses [23], has shown that the viscoelastic moduli of shear induced crystals have smaller values than the amorphous/glass of the same φ, although qualitatively similar frequency behaviour. However, when compared to the same distance from maximum packing, the ordered structure exhibits larger elastic moduli. The yield strains of the crystal were additionally observed to be smaller than those of their counterpart amorphous.

Light scattering experiments [38, 40] showed the structures and geometric flow characteristics of the hard sphere crystal under shear. Other work studied colloidal crystallization with a confocal microscope using a counter rotating cone plate shear cell and was able to image the shear behavior of hard sphere crystals by looking at a stationary plane[52]. Moreover, confocal microscopy was used to measure the mean squared displacements of a crystal under steady shear [53], noting the zig-zag and collective layer motions of a crystal, while discussing the crystal melting with a dynamic Lindemann criterion and the measured long and short time shear induced diffusion coefficients. Later work with confocal microscopy experiments on crystallization and melting under shear discovered that crystal melting occurs in domains and that crystallization under shear is not based on nucleation[54]. The differences in the microscopic motions of glass and crystal under shear however have not be thoroughly examined, while there is interest in understanding the interplay of applied oscillation frequency and strain amplitude during yielding.

In this work we have implemented Brownian dynamics simulations on shear induced hard sphere crystal structures under oscillatory shear and make comparisons to their amorphous counterparts. The purpose of this work is the detailed determination of the microscopic rearrangements under oscillatory shear and their relation to mechanical properties for the purpose of understanding the mechanisms behind shear-induced crystallization. For this purpose we performed BD simulation in polydisperse glassy configurations and FCC crystals with close packing parallel or perpendicular to shear.

## Simulation method and analyses

Brownian Dynamics (BD) is a simulation method which treats the fluid as a continuum, but particles are still small enough to be influenced by Brownian fluctuations. BD can be described as a simplification of Stokesian dynamics [55], where now hydrodynamic interactions between particles are neglected. For N rigid particles of radius R and density ρ in a medium of viscosity η moving with velocity **U**, we examine states where the Reynolds number is $Re<<1$ (the dimensionless ratio of inertial forces $\rho U^2 R$ to viscous forces $\eta U$). In this case the motion of the particles is described by the N-body Langevin equation, $\mathbf{m}(d\mathbf{U}/dt) = \mathbf{F}^H + \mathbf{F}^B + \mathbf{F}^P$, where **m** is the generalized mass/moment tensor, **U** is the particle translational/rotational velocity vector, $\mathbf{F}^H$ is the hydrodynamic force vector, $\mathbf{F}^B$ is the stochastic force vector that gives rise to Brownian motion, and $\mathbf{F}^P$ is the deterministic non-hydrodynamic force vector. Since inertia is not important in colloidal dispersions (Re<<1) the equation reduces to: $0 = \mathbf{F}^H + \mathbf{F}^B + \mathbf{F}^P$. Since we assume negligible hydrodynamic interactions between particles, the hydrodynamic force reduces to Stokes drag $\mathbf{F}^H = -6\pi\eta R\mathbf{U}$. The non-hydrodynamic force vector in these simulations is only related to the hard sphere interaction occurring at contact between particles.

The interactions between hard sphere particles are calculated based on a "potential-free" algorithm [56], in which the overlap between pairs of particles is corrected by moving the particles with equal force along the line of centres, back to contact. This algorithm is "potential free" concerning the hard repulsion in that it does not require a specific declaration of a pair potential, although it implements the hard-sphere potential which is infinite if the particles are overlapping and zero otherwise. In order to calculate the stress, the algorithm directly calculates the pairwise interparticle forces that would have resulted in the hard sphere displacements during the course of a time step [57]. Therefore, $\mathbf{F}^P = 6\pi\eta R\left(\Delta x^{HS}/\Delta t\right)$, the average Stokes drag on the particle during the course of the hard-sphere displacement. The interparticle forces then can be used to calculate the stress tensor, $\langle\mathbf{\Sigma}\rangle = -n\langle\mathbf{xF}^P\rangle$ [57], where $n$ is the number density of particles and the angle brackets denote an average over all particles in the simulation cell. The position and displacements of the particles are non-dimensionalized by the particle radius $R$, and the time $t$ by the Brownian time $t_B=R^2/D$.

In the case of a simple steady shear we could define a non-dimensional shear rate from the Peclet number, $Pe = \dot{\gamma}\cdot t_B = \dot{\gamma}\cdot R^2/D$, specifically meaning that for $Pe=1$, 100% strain will be reached in a time scale of $t_B$. For the case of oscillatory shear a different non-dimensional number must be used. We define $Pe_\omega$, which corresponds to an oscillation frequency with $\gamma = \gamma_0 \cdot \sin(Pe_\omega \cdot t)$ and $Pe_\omega = \omega R^2/D = 2\pi t_B/T$, where $\gamma_0$ is the peak amplitude of shear and $\omega$ is the frequency of oscillation and $T$ is the period. We may compare the maximum Peclet number in the oscillation with steady rate shear using the relation given by $Pe_{max}=\gamma_0 Pe_\omega$. As a final note, BD simulations have been found to qualitatively capture experimental stresses and particle motions at rest and for relatively low $Pe$, even with the absence of hydrodynamics [55].

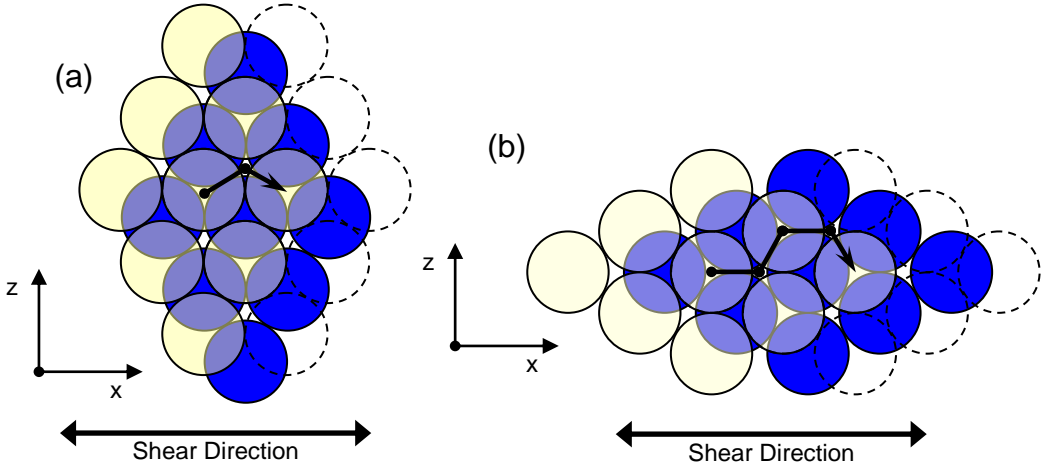

Figure 1: Schemes showing a) the simple slipping layer yielding process of a crystal with the close packed direction parallel to shear and b) the complex slipping layer yielding process of a crystal with the close packed direction perpendicular to shear.

In the experimental systems, the crystals where created by the application of shear, however initial testing with simulations showed that crystallization in a finite simulation box creates crystals which are skewed and randomly oriented, even under shear. This occurs simply because the propagation of the crystal due to Brownian motion in the simulation cell occurs faster than shear can align it. As a second point the anisotropic nature of the FCC crystal in the shear induced orientations requires that the periodic box must be of dimensions tailored to fit multiples of the FCC crystal unit cell.

Therefore, the simulated hard sphere FCC crystals were initially constructed to the appropriate specifications and then left to reach equilibrium at rest for t=5$t_B$ before applying shear. Depending on the simulation, the (111) plane was positioned on the velocity-vorticity plane with the closed packed direction of the spheres set either parallel (fig. 1a) or perpendicular (fig. 1b) to the shear direction. As shown in the Supplementary Information (SI), there is a dependence of the simulation results on the anisotropy of the simulated box. In general it was found that isotropic boxes are more suitable for looking at systems at rest, while the highly anisotropic ones for looking at crystal systems under shear and the particulars of the yielding. Thus, two types of simulated boxes for the crystal are shown in this paper, one relatively isotropic with 1440 particles and one highly anisotropic (longer in the velocity gradient direction) with 4860 particles. We denote them as 10-12-12 and 6-135-6, representing the number of particle layers in the velocity (x), velocity gradient (y) and vorticity (z) directions respectively. In order to better replicate the experimental conditions, the 6-135-6 system has a random FCC layering in the y-axis.

The amorphous polydisperse glass was simulated in a cubic box with 1440 or 1005 particles both at rest and under shear and a Gaussian distribution of radii with 10% standard deviation to suppress crystallization. The particles in both the amorphous and crystal system were randomly placed and left for some time (a maximum of 300$t_B$) to reach a equilibrium before applying shear. Simulations were run with a time step $\Delta t$ of $2\pi$ $10^{-4}$ $t_B$ or less and displacements under shear were averaged over at least 15 oscillations.

Analysis of particle motions is done through the Mean Squared Displacement (MSD), which gives a statistical measure of the distance a particle has moved in a specific time. If $x_i$ is the non-affine position (by subtraction of affine shear field motion) of a particle $i$ in the $x$ direction and $N$ is the total number of particles then the MSD is generally calculated by using the following equations:

$$\left\langle \Delta x^2(\tau) \right\rangle = < \frac{1}{N} \sum_{i=1}^{N} [x_i(t+\tau) - x_i(t)]^2 >_t \text{ and}$$
$$\left\langle \Delta r^2(\tau) \right\rangle = \left\langle \Delta x^2(\tau) \right\rangle + \left\langle \Delta y^2(\tau) \right\rangle + \left\langle \Delta z^2(\tau) \right\rangle$$
(Eq. 1)

where $x_i$ is substituted with $y_i$ and $z_i$ for the different axis. However, because of the finite system size, the high density of the system and subsequent small displacements over long times, subtraction of the center of mass motion is needed as a correction, such that:

$$\left\langle \Delta x^2 \right\rangle_t = < \frac{1}{N} \sum_{i=1}^{N} \left[ (x_i(t+\tau) - x_i(t)) - \frac{1}{N} \sum_{i=1}^{N} [x_i(t+\tau) - x_i(t)] \right]^2 >_t \text{ (Eq. 2)}$$

It is important to note that by subtracting the affine motion at every time step, the effect of Taylor dispersion is lost [58].

The one sided, self Van Hove function [59], $G_S(\Delta x, \tau)$, is the probability density function of a particle's displacement for an elapsed time $\tau$:

$$G(\Delta x, \tau) = \frac{1}{N}\left\langle \sum_{i=1}^{N} \delta\left(\Delta x - |x_i(t+\tau) - x_i(t)|\right) \right\rangle_t = \frac{N(\Delta x, \tau)}{N} \quad \text{(Eq. 3)}$$

where Δx is the distance a particle has moved in a time scale τ, averaged over time. For a single sphere in a Newtonian fluid, this Van Hove function has a Gaussian profile. The lowest order deviation of the Van Hove function from Gaussian behavior is given by:

$$a_{2x}(\tau) = \frac{\langle \Delta x^4(\tau) \rangle_{N,t}}{3\langle \Delta x^2(\tau) \rangle_{N,t}^2} - 1 \quad \text{(Eq. 4)}$$

averaged over particles and time, where $a_{2x}$ is calculated for a single dimension. Non-zero values mean that $G_s$ exhibits non-Gaussian behavior and have been associated with dynamic heterogeneities [60], although non-Gaussian behavior may not always be related to dynamic heterogeneities.

## Glass vs. Crystal at Rest

We start off by comparing the displacements of the FCC crystal to a glass without the application of shear. Figure 2 shows the mean squared displacements (MSD) of the simulated crystal and glass at rest for various volume fractions both below and above the glass transition. Since we are examining a polydisperse system, the arrest transition is expected to increase above 0.58[61]. The crystal shows an initial increase of the MSD reflecting an in-cage beta relaxation and then a plateau. The amorphous configurations show an initial increase (beta), a region of smaller slope and then depending on φ, a strong increase of the MSD or weaker increase. The former corresponds to lower φ out of cage alpha relaxation, while the latter suggests the suppression of out of cage diffusion (between φ=0.60-0.62), with the existence of some dynamic heterogeneities [60, 61]. Depending on the time scale of probing and the volume fraction, the amorphous structure may have a higher MSD than the crystal at long times, but less at short times. Both the amorphous and the crystal generally have isotropic displacements at rest which increase in amplitude as φ is decreased.

Generally, as φ is increased, the available volume for each particle within its cage of neighbors decreases, causing a general drop of short time particle displacements. We find that at high volume fractions, where there is small or no alpha relaxation, the MSD in a glass is always less than that of the crystal. This occurs because the average available free volume for each particle is always less in the glass when compared to the crystal of the same volume fraction [23]. As the volume fraction is decreased though, dynamic heterogeneities and the α-relaxation become important, so the amorphous configurations exhibit larger MSDs at longer times than the ordered and stable crystal.

As mentioned, particles undergoing random diffusive motion should show displacement distribution profiles, or self Van Hove functions, which show Gaussian statistics. However in the case of glassy suspensions, the particles are not diffusive for all time scales, thus we expect deviations from Gaussian behavior. Such deviations in these systems has been attributed to dynamic heterogeneities [6, 7, 22], which occur when clustered groups of particles move at different speeds than the bulk. Dynamic heterogeneities then refer to two distinct populations of particles, those that follow motions mainly within the cages and those which undergo faster out of cage jumps due to cooperative motions. We note that since the displacements for the glasses and liquids at rest are isotropic, the findings are equivalent in all directions.

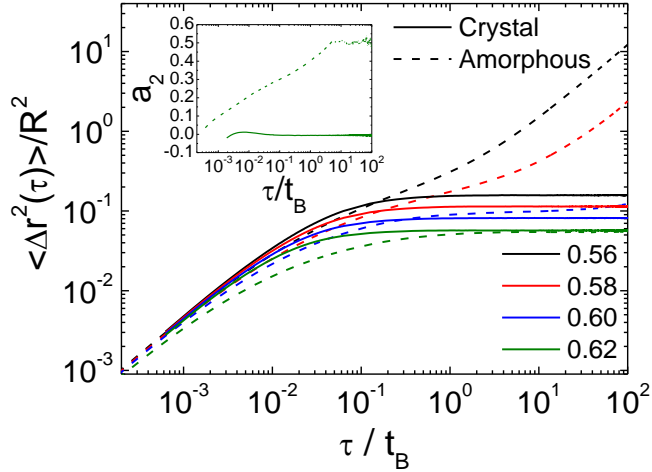

Figure 2: Mean squared displacements of amorphous polydisperse configurations and FCC crystal configurations (10-12-12) at rest for different $\varphi$. Inset shows the non-Gaussian parameter, $a_2$, for $\varphi=0.62$ at rest for the crystal and the glass.

We may quantify the deviation from Gaussian statistics by calculating the $a_2$ parameter as shown in the inset of figure 2, at $\varphi=0.62$ for the crystal and the glass. In the case of a system at rest, the parameter gives a measure of the dynamic heterogeneities present in the simulated system. For the glass, heterogeneities are low at the smallest time scales since particles move diffusively in their cages. As the probed time is increased however, heterogeneities increase as correlated motions arise. The crystal on the other hand, consistently stays at a value of near zero, showing constant Gaussian-like behavior. Unlike the dynamic heterogeneities found in the metastable glass, the displacements of the crystal particles at rest are spatially homogeneous.

## Glass vs. Parallel Crystal under Oscillatory Shear
### Glass vs Crystal under Shear: Mean Squared Displacements

After examining and comparing the dynamics at rest for the crystal and glass, we move on to study the occurring phenomenology for oscillatory shear. In figure 3a we show the MSD of a glass and a crystal parallel to shear at $\varphi=0.62$ as a function of time for a specific strain of 30% and $Pe_\omega=1$. In the case of the glass, the displacements are isotropic, with the $<\Delta x^2>$, $<\Delta y^2>$ and $<\Delta z^2>$ curves being approximately equal. In the case of the crystal however, the curves follow different paths showing highly anisotropic behavior. Displacements in all directions are approximately the same at short times, corresponding to in-cage motion, but at longer times, where out of cage motion due to shear occurs, $<\Delta x^2>$ (velocity) exhibits the highest increase, followed by $\langle \Delta z^2 \rangle$ (vorticity) and lastly $\langle \Delta y^2 \rangle$ (velocity gradient), which remains almost stable.

Figure 3b depicts the MSD for the different directions with respect to shear at a specific time equal to one period ($t=T=2\pi t_B$) for the $\varphi=0.62$ glass and crystal. Again the anisotropy of the crystal under shear is clear when compared to the glass. At strains greater than 5%, the beginning of yielding [23], the particle displacements in the glass initially show a sharp and then a weaker increase with strain. Similarly the crystal shows substantial changes above the yield strain. The greatest increase is in the x direction and persists up to 100%. However the MSD in the z-axis, slowly decreases with increasing strain after the initial jump. The y-axis displacements do not show abrupt changes around the yield point, but exhibit a slight decrease at higher strains. The inset depicts the average displacement behavior ($\langle \Delta r^2 \rangle$) in the glass and the crystal with applied strain again at $\tau=T$. The apparent finding from figure 3b is that while the glass and crystal exhibit similar overall displacements in the linear regime (see also fig 2), at strains above the yield the glass has significantly larger displacements which furthermore increase faster with strain than the crystal.

The glass in fig. 3a qualitatively shows a much different behavior than the crystal. The short times which correspond to in-cage motion, exhibit weaker displacements than the crystal, as explained by the decreased free volume per particle [23], similar to the situation at rest. At longer times, which are comparable to the period of oscillation, shear induced out of cage displacements occur showing a diffusive character [36]. The similar values of long time displacements for the crystal in the x-axis and for the glass are coincidental as also can be seen in the strain dependence of fig. 3b.

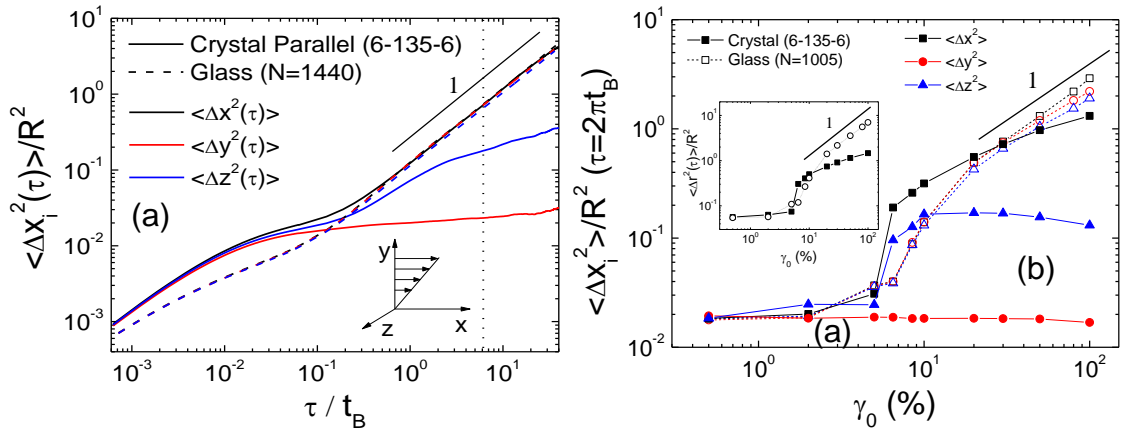

Figure 3: Mean square displacements of a glass and crystal parallel to shear at $\varphi=0.62$ under oscillatory shear at $Pe_\omega=1$ with a) the MSD at different directions vs time at 30% strain and b) the MSD at different directions vs Strain at $\tau=T=2\pi t_B$ and overall displacements vs Strain at $\tau=T=2\pi t_B$ (inset). The vertical dotted line in a) refers to the time scale of points in b), while the solid lines in both a) and b) are power law slopes of 1.

By looking at a specific time scale, in this case the period of oscillation (τ=T), we may examine the MSDs in relation to the maximum strain $\gamma_0$ and thus the shear rate $\dot{\gamma}$ (figure 3b). The displacements in the glass start with low values at low strains, reflecting the low in-cage plateau. The sharp increase above the yield strain (>5%) with an exponent of greater than 1, is related to the glass cage that starts to "melt" as particles start to diffuse out of their cage under the influence of shear (the $<\Delta r^2>$ vs τ exponent is approaching unity at long times). In the glass, diffusive behavior is observed in this time scale for values of $\gamma_0$>10%. The linear increase ($<\Delta r^2>$ vs $\gamma_0$ exponent of roughly 1) seen at higher strains (>10%) can be ascribed to the $\gamma_0$ and $\dot{\gamma}$ increase of diffusivity [27, 57, 62] and has been found to be associated to rheology with a power law slope, which depends on the frequency and volume fraction [26, 36].

With the help of the scheme of fig. 1a, which depicts the crystal layers and sliding mechanism and in agreement with previous findings and suggestions [40, 48, 53], one can explain the MSD findings in figures 3a and 3b. In the glass, particles have random isotropic positions, so at rest and under shear, the displacements in the x, y and z directions are almost isotropic [27, 33, 36, 57, 62]. In the case of the crystal, however, even though the positional anisotropy of the particles does not cause any strong anisotropic displacements at rest, it shows strong anisotropy under shear. A movie of the unbounded motion of the particles under shear in both the glass and the crystal is provided in the SI.

As shown in the scheme of fig. 1a, in a crystal, the particles under shear are confined to their crystal layers in the x-z plane. Shear causes the particle layers to slide over each other, giving rise to the displacements seen in fig. 3a and 3b. The figures show that at strains above yield, the largest displacements in the crystal are in the x direction (velocity), as the layers slide in this direction. The z direction (vorticity) has the second largest displacements, due to the layer zig-zag motion that allows the sliding to take place. The confinement in the x-z plane means that the smallest displacements occur in the y direction (velocity gradient), which can be mostly attributed to in-cage diffusion. In fig. 3a, $<\Delta x^2>$ shows diffusive behavior for long times, reaching a power law of one, though it should be clarified that such diffusion occurs through the motion of layers as whole rather than individual particles as will be shown below. The out of cage $<\Delta z^2>$ shows a strong initial increase directly related to the zig-zag motion, while the weaker long time increase should be related to a longer time scale layer hopping. Finally, we note that the increase of $<\Delta y^2>$ at longer times is due to an artifact of the large anisotropy of the box and is discussed in the SI.

While having isotropic displacements at low strains in fig. 3b, the crystal exhibits high anisotropy above the yield strain (>5%). The increase in $<\Delta x^2>$ above the yield strain occurs, as discussed, because of increased displacements due to sliding layers. Above the yield strain there is also a sudden increase of $<\Delta z^2>$, from the zig-zag motion of the layers, although the values dwindle with increasing strain as higher shear rates constrict the motion to specific pathways lowering the in-cage Brownian motion. Similarly, $<\Delta y^2>$ slightly decreases as the higher rate sliding layer motion constricts random Brownian motion in the y-axis as well.

The overall displacements in the inset of fig. 3b show that although the MSD of the glass and crystal are comparable in the linear regime, the glass shows significantly more long time displacements in the nonlinear regime than the crystal. At $\gamma_0>10\%$, particles in a glass rearrange a diffusively at the time scale of one period and $<\Delta r^2>$ vs $\tau$ shows close to a linear increase. At $\gamma_0 > 20\%$ particles in a crystal rearrange anisotropically with average displacements always less than in the glass. Contrary to the glass, a peculiar finding is that both the exponents of $<\Delta r^2>$ vs $\gamma_0$ and also $<\Delta x^2>$ vs $\gamma_0$ are about 0.6.

Experimental findings of [53] on a crystal under steady shear showed a linear increase of both long and short time diffusivity. This is not the case for our findings as we extract a sub-linear slope of 0.6 for long time displacements, while short time displacements related to the in-cage diffusivity show almost no changes when compared to rest. The longtime diffusivities were examined in a small range of shear rates, which may have given an apparent linear dependence with rate. Moreover, their experiments were carried out under the application of steady shear, which may qualitatively change particle motions in comparison to oscillatory shear.

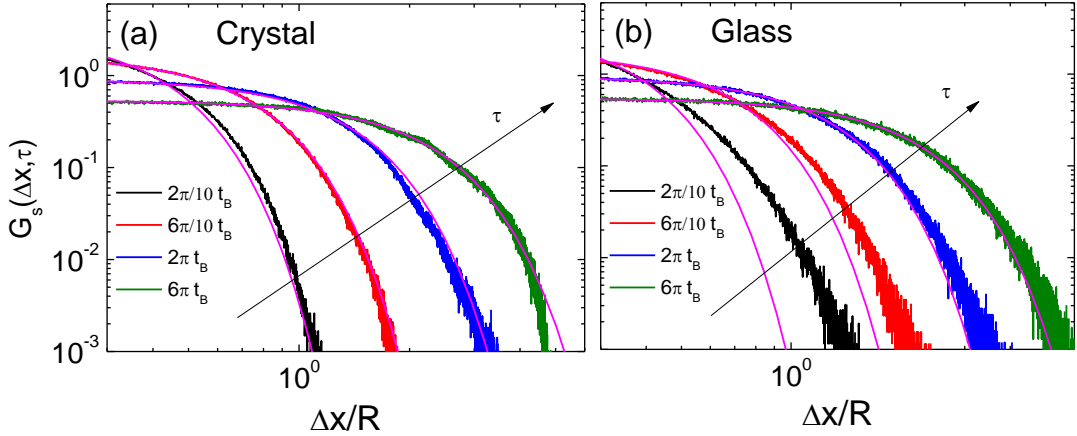

Figure 4: x-axis single sided Van Hove self-correlation functions (probability density function) at φ=0.62 under oscillatory shear at $Pe_\omega=1$ and 30% strain amplitude at various time scales $\tau/t_B=2\pi/10$, $6\pi/10$, $2\pi$ and $6\pi$ with $T=2\pi$ for a) the crystal parallel to shear and b) the glass. Magenta lines are Gaussian fits to the data.

The higher MSD values in the crystal around $\gamma_0=10\%$ can be explained by the difference of yield strains between crystal and glass. The onset of yield in the glass shows a smaller strain than the in crystal, although the crystal yields more abruptly in a smaller range of strains due to the slipping layers, thus the particle displacements in the crystal are higher than in the glass around the crystal yield strain. This is seen in the rheological measurements of [23], with the onset of yield being earlier for the glass, while reversely the solid to liquid transition occurring sooner for the crystal. Such behavior may not be observed if we probe a higher volume fraction glass with a smaller yield strain. Additionally, it is important to note that as the probed time scale decreases, thus monitoring in-cage motion, the qualitative differences between a glass and crystal should also decrease, as may be inferred from figure 3a. Generally we find that the average long time particle displacements in the crystal under shear are smaller than in the glass of the same volume fraction. At shorter times though, which reflect the in-cage motions, particle displacements in the crystal are larger than those in the glass due to the increased free space per particle.

**Glass vs Crystal under Shear: Non-Gaussian Behavior**

In order to quantify the deviation of the displacements under shear from Gaussian statistics we plot Van Hove probability density functions along with the non-Gaussian parameter. We focus our analysis on the velocity direction, where the most significant motion occurs for the crystal. Fig. 4 shows the x-axis Van Hove functions of the crystal (4a) and glass (4b) at $Pe_\omega=1$ and $\gamma_0=30\%$ as seen for the MSD in fig. 3a with the corresponding Gaussian fits. The time scales probed range from within the period (at T/10) to a multiple of the period (at 3T). The crystal roughly shows Gaussian behavior for all time scales, while the glass shows stronger deviations as time scales become smaller within the period. Summarizing these findings, figure 5 shows the $a_2$ parameter as a function of time for the crystal and glass under shear again at $Pe_\omega=1$ and $\gamma_0=30\%$, also comparing to rest. The crystal under shear shows similar values to rest for the smallest times, while for larger times, although still before the time scale of a period, there is a peak. Time scales larger than a period fluctuate towards a zero value. The glass under shear, exhibits higher values of $a_2$ than at rest for small times, which increase and show a peak for time scales less than an oscillation period. At times larger than a period, the glass particle displacements become Gaussian.

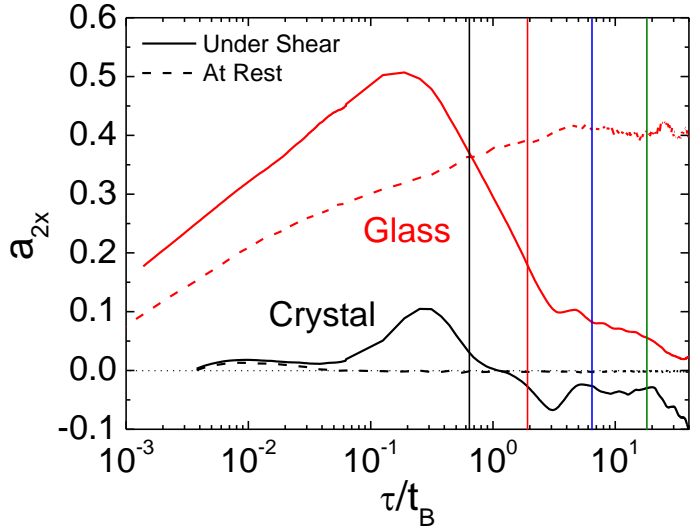

Figure 5: Non-Gaussian parameter in the x-axis versus time for the crystal parallel to shear (black) and the glass (red) at $\varphi=0.62$ for $Pe_\omega=1$ and $\gamma_0=30\%$ as in fig. 3a and at rest. The vertical lines correspond to the time scales probed in figures 4a and 4b.

The slope of unity for the x axis displacements of the crystal (fig. 3a) at long (and short) times together with behavior of the Van Hove function (fig. 4a) and the $a_2$ parameter (fig. 5) suggests that particle displacements are Gaussian at these time scales. For the glass (fig. 3a) the displacements at long time increase linearly with time, while $a_2$ is close to zero, suggesting diffusive out of cage motion due to shear. However at shorter times the displacements are non-Gaussian. In figure 5, the non-Gaussian parameter of the glass under shear increases with time prior to flow, because of contributions of dynamic heterogeneities at rest, as well as the non-Gaussian contribution from particle yielding due to shear, dropping to zero, when out of cage diffusivity becomes important at long times. Similarly, the crystal with minimal non-Gaussian behavior at rest shows a peak at the time scale that marks the transition from the in-cage diffusion to the shear-induced sliding layer diffusion, and resuming Gaussian behavior at long time scales.

## **Cooperative Motions of Crystal under Shear**

To investigate how cooperative particle displacements occur in a sheared crystal, we calculate the mean squared displacements of each particle in the crystal after subtraction of the motions of their x-z plane neighbors. In order to do this we modify the equation for the calculation of the MSD, Eq. 2 as:

$$\left\langle \Delta x_{neib}^2 \right\rangle_t = < \frac{1}{N} \sum_{i=1}^{N} \left[ (x_i(t+\tau) - x_i(t)) - \frac{1}{Z} \sum_{k(i)}^{Z} \left[ x_{k(i)}(t+\tau) - x_{k(i)}(t) \right] \right]^2 >_t \quad \text{(Eq. 5)}$$

where Z is the total number of neighbor particles. In this case, Z=6 and the *k* particles are identified as the closest neighboring particles of the same x-z layer at the beginning of the simulation, remaining the same throughout the calculation. Specifically, a Z number of *k* particles are chosen from the initial configuration when satisfying these conditions in relation to the *i* particles: $|\bar{r}_i - \bar{r}_k|$ : min and $|y_i - y_k|$ : min.

Figure 6 shows particle displacements under shear relative to their x-z layer cage neighbors for the crystal parallel to shear at φ=0.62, between $\gamma_0$=2% and 100% strains with $Pe_\omega$=1. Displacements in the velocity gradient direction (y) have not been plotted, since $<\Delta y_{neib}^2>$ is statistically similar to $<\Delta y^2>$, as the single particle motion in this axis is uncorrelated to its neighbors (for this φ). The relative displacements seem to remain practically constant at all times under shear, except in the case of $<\Delta x_{neib}^2>$ for 100% where an increase at longer times is observed in the linear regime ($\gamma_0$=2%) and similarly to the displacements at rest, $<\Delta x_{neib}^2> \approx <\Delta z_{neib}^2>$. In the non-linear regime, at $\gamma_0 \geq 10\%$, $<\Delta x_{neib}^2>$ steadily increases with increasing strain, while $<\Delta z_{neib}^2>$, decreases. The Van Hove functions for $\gamma_0$=30% for the x-axis displacements at various time scales shown in the inset are practically identical and of Gaussian character.

The relatively constant values of $<\Delta x_{neib}^2>$ and $<\Delta z_{neib}^2>$ under shear compared to their respective $<\Delta x^2>$ and $<\Delta z^2>$ show that indeed the previously discussed displacements occur as collective motions of whole x-z layers and do not correspond to individual particle motion. This also means that the values of $<\Delta x_{neib}^2>$ and $<\Delta z_{neib}^2>$ only reflect the changes in the in-cage Brownian motion of the particles, since particles are trapped in their layers.

Consequently, the decrease of $<\Delta x_{neib}^2>$ or $<\Delta z_{neib}^2>$ corresponds to a constriction of the cage in the specific direction, while an increase suggests an elongation. Hence fig. 6 indicates a constriction of the cage with increasing strain in the z-axis, in addition to that seen before in the y-axis (fig 3b).

The increase of $<\Delta x_{neib}^2>$ at high strain and longer times points to an elongation, which is possibly an inlayer cage stretching due to strong collisions. More accurately it hints towards a fluctuating cage distortion, instead of a simple static elongation, which requires more than a single period of oscillation for a particle to probe its length. The alternative of out-of-cage displacements within the layer, has been ruled out by neighbor enumeration throughout the analysis, which shows that no overtaking events occur. Assuming this increase is due to inlayer cage stretching, in the limit of infinite time, relative displacements, $<\Delta x_{neib}^2>$, should reach a second plateau. This has not been directly seen, but there are some indications of a plateau value for applied strains at around 10%. Unfortunately, verification of this would generally require simulated time scales of a few orders of magnitude more than this work.

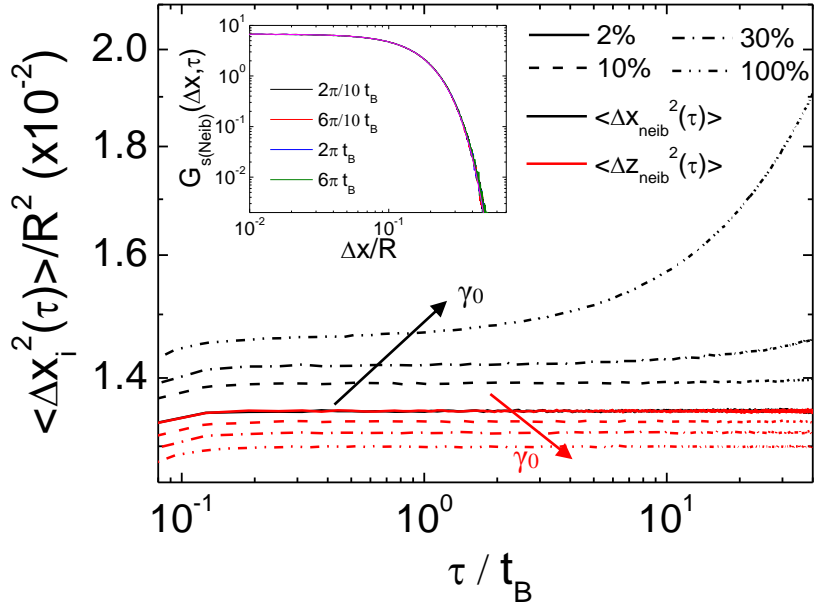

Figure 6: Mean squared displacements under shear from a crystal parallel to shear at $\varphi=0.62$ and $Pe_\omega=1$ for $\gamma_0=2\%$, 10%, 30% and 100% strain amplitudes. Values of $<\Delta y_{neib}^2(\tau)>$ have been omitted as they are statistically equal to $<\Delta y^2(\tau)>$. The inset shows the Van Hove function of the relative displacements for $\gamma_0=30\%$ in the x axis at different times with a Gaussian fit (magenta), showing a complete overlap.

Additionally we find that for these time scales, the displacements are quite far from the Lindemann criterion of crystal melting[63], which would need $\sqrt{\langle \Delta x_{neib}^2 \rangle} > 0.26R$, whereas the maximum value shown in figure 6 is $\sqrt{\langle \Delta x_{neib}^2 \rangle} = 0.14R$. Indeed from the state at rest of figure 2, the limit of $\varphi=0.545$, where the crystal is expected to melt, we may estimate $\sqrt{\langle \Delta x_{neib}^2 \rangle} = 0.26R$. Moreover, as seen from the Van Hove function inset, the relative crystal displacements in the x-axis are Gaussian for $\gamma_0=30\%$. The rest of the different strains and axis show similar Gaussian behavior at all shown time scales.

## **Glass vs Crystal under Shear: Dynamics and Stresses within the Period**

The particle MSDs under oscillatory shear examined up to this point, were averaged between configurations throughout the period. This may be expected to obscure details of the particle motion, since the sheared particles are not at a steady state and shear rates are not constant. A next step would be to look into the particle motions under shear within the oscillation period and additionally examine their correlation to shear stress. Figure 7 shows the instantaneous MSD, $<\Delta x_i^2>_{inst}$, and shear stresses of the crystal and the glass for $\varphi=0.62$, $\gamma_0=30\%$ and $Pe_\omega=1$, in order to understand the rearrangements within the period of oscillation in the non-linear regime. $<\Delta x_i^2>_{inst}$ are determined by averaging displacements over a small part of the period ($\tau=2\pi/100=T/100$) and can be thought as an averaged particle rate of motion.

In both the cases of crystal parallel to shear (fig. 7a) and amorphous glass (fig. 7b) the displacements are quite complex and change within the period of oscillation, contrary to the more trivial case of the linear regime ($\gamma_0<2\%$ - not shown) which remain constant. Due to anisotropy and layer displacements, the crystal does not exhibit stresses and displacements that are simply proportional to the rate, but seem to depend on both the elapsed strain and the rate (fig. 7a). When the rate is zero and the strain is maximum at $t/t_B=\pi/2$, we expect the crystal to behave similarly as at rest (for relatively low $Pe_\omega$), as the small rate allows the dynamics to relax from previous shear history, although keeping the elastic stress from previous deformation. When the rate approaches the maximum at $t/t_B=\pi$, the stress changes sign and approaches a peak which coincides to strong changes in the dynamics, a maximum in $<\Delta x_i^2>_{inst}$ and a minimum in $<\Delta z_i^2>_{inst}$. After this point, there are some undulations of the stress and then a reversal of the process at zero shear rate ($t/t_B=3\pi/2$). As seen in previous experimental work, the crystal generally has less stresses than glass in the nonlinear regime[23].

The glass displacements in fig. 7b are somewhat simpler than the crystal following the rate of shearing and the stresses more closely. We find isotropic displacements and stresses which are roughly proportional to the rate. When the isotropic glass is sheared in the non-linear regime, the system is shear melted and flows with particles forced out of their cages. The proportionality of the stress to the shear rate shows viscous behavior and thus the increased displacements under shear correspond to the energy dissipation related to viscous stresses and out of cage displacements[36]. The complex behavior of the crystal within the period can be explained by the anisotropy and sliding layer yielding process. As described, there is a peak in the stress response approximately after 20% strain has elapsed from the point of zero shear rate. The x-axis instantaneous displacements show a slightly delayed, but sharp increase, along with the occurrence of the peak of the strain. Thus as the motion in the x-axis corresponds to layer sliding, we conclude that the stress peak corresponds to the storage of energy as the system is strained and then the release of this energy when the layers begin to slide. The y-axis shows a minimum instead of a peak, due to the constriction that occurs during the sliding. The undulations after the first peak are caused by subsequent sliding of the layers, which are present only in the crystal because of the reproducibility of the structure with increasing strain and that should not be apparent in the amorphous glass.

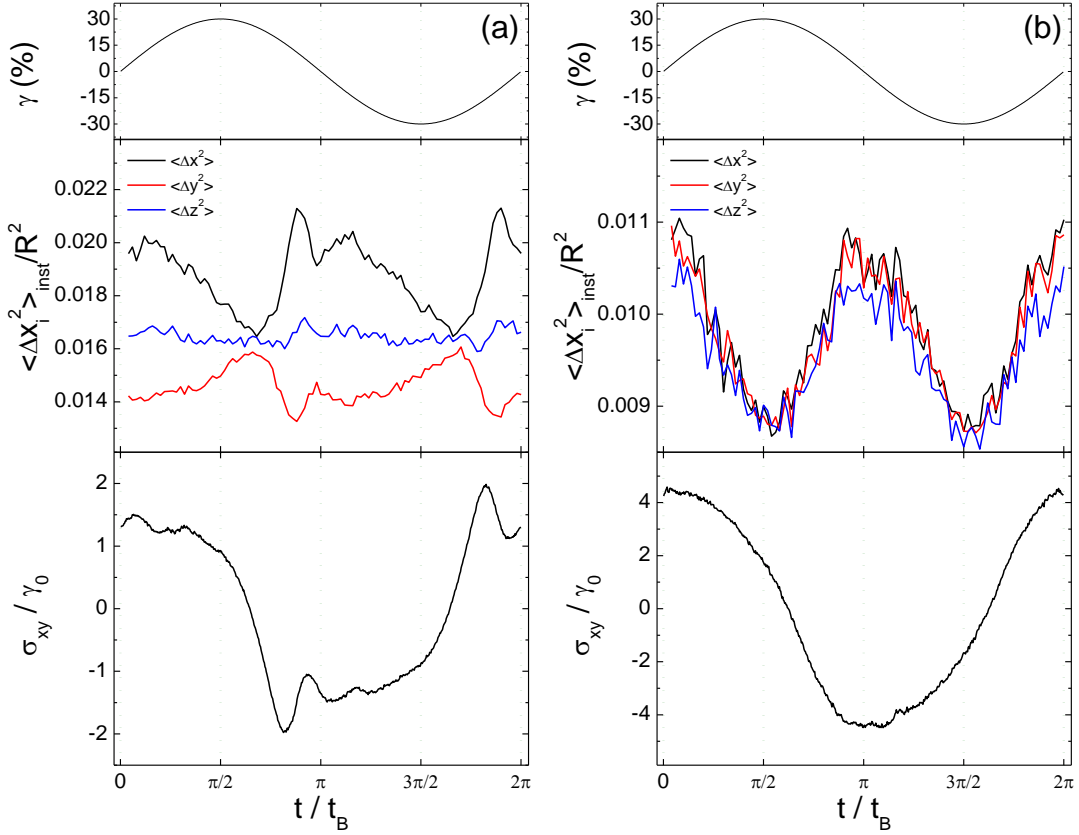

Figure 7: Mean squared displacements at $\tau=2\pi t_B/100$ ("instantaneous" MSD) for each axis at $\varphi=0.62$, $Pe_\omega=1$ and $\gamma_0=30\%$ within the period (top) and normalized stress (bottom) versus time of a) the crystal parallel to shear and b) the same $\varphi$ glass

The instantaneous particle displacements within the period in the crystal, on average, are larger than in the glass, while reversely the stresses in the glass are larger than in the crystal. This occurs because the measured stress does not correspond to the absolute values of the instantaneous MSD; rather they roughly correspond to the change of displacements from the random Brownian motion average at that time scale. Since $<\Delta x \Delta y>$ at rest is zero and the displacements in the x and y directions are uncorrelated, we are able to directly get $\sigma_{xy} \sim <\Delta x \Delta y>_{(\tau \to 0)}$ from the simulation method. However, $\sigma_{xy}$ is generally is not directly relatable to $<\Delta x^2>$, $<\Delta y^2>$ or $<\Delta z^2>$ as random Brownian motion also contributes to these values.

## **Glass vs Crystal under Shear: Discussion**

Another way of examining the properties of the crystal and the glass and quantitatively comparing the viscoelastic properties with the MSD is using the Generalized Stokes Einstein (GSE) equation from [64]. The GSE relates the complex shear modulus with mean square displacements as $G^*(\omega=2\pi/\tau) \propto 1/<\Delta r^2(\tau)>$. Thus the plateau value of the mean square displacement, quantifying the localization of the particle should be inversely proportional to the elastic modulus. However we observe that at $Pe_\omega=1$ (or $\tau=2\pi$) and linear strains for $\varphi=0.62$, the isotropic plateau MSD, for the glass and the crystal is comparable (as in fig. 2), but elastic modulus of glass is larger than that of crystal. Therefore in contradiction to the GSE relation from [64], while the crystal and the glass in simulations have similar MSD, the moduli are quite different. In this formulation, it may be that the changes in the structural properties (i.e. amorphous vs ordered) and thus the elastic constants are not taken into account, or that the polydispersity of the glass is affecting the dynamics. We would like to add however, that as the glass approaches maximum packing $\varphi_{RCP} \approx 0.64$ [65] and particles reach the limit of immobility, the corresponding FCC will be far from the maximum crystal packing $\varphi_{FCC} \approx 0.74$ and thus still quite mobile.

Moreover, the FCC crystal is the entropically favorable configuration of concentrated hard sphere dispersions [16, 66]. When in the glass regime, the particles find themselves kinetically trapped and unable to rearrange into the minimum free energy state. The application of shear provides the particles in the glass regime the ability to rearrange and explore the energy landscape with large and isotropic MSD as seen in fig. 3. Eventually the particles find the minimum free energy (maximum of entropy) and as seen in experimental studies [23, 38-41], the system crystallizes with specific orientation compared to shear. As seen in figs. 3 and 7, the crystal that is created under shear has both less shear stress and fewer displacements which are additionally anisotropic.

## Parallel Crystal under Oscillatory Shear

### Crystal under Shear: Frequency Dependence

Now turning our focus solely to the crystal parallel to shear, we move to examine the frequency response under nonlinear oscillatory shear. Figure 8a shows the strain dependence of the MSD for two different frequencies, $Pe_\omega=0.1$ and $Pe_\omega=1$, at $\varphi=0.62$ separately for the three axis of motion, while figure 8b shows the x-axis for $Pe_\omega=0.1$, 1 and 10. The extracted MSDs are at a time scale of one period ($20\pi$ $t_B$ for $Pe_\omega=0.1$, $2\pi$ $t_B$ for $Pe_\omega=1$ and $2\pi/10$ $t_B$ for $Pe_\omega=10$). For $Pe_\omega=1$ the data in fig. 8 is the same as plotted and described for fig. 3b. The curves for all directions and $Pe_\omega=0.1$ in 8a have larger values than for $Pe_\omega=1$ and the onset strain of nonlinearities is smaller for $Pe_\omega=0.1$ (>2%). Additionally, between the two frequencies, the displacements have qualitatively similar shapes. Similarly to 8b, displacements increase with decreasing frequency and again nonlinear behavior begins at lower strains for lower frequencies.

We find that the increase of the displacements at lower $Pe_\omega$ is due to the larger time scale per period that allows Brownian motion to enhance the shear induced displacements during sliding. This is also seen experimentally in glasses under oscillatory shear [25] as well as in simulations[36]. Looking at both 8a and 8b, it seems that the yield strain for lower $Pe_\omega$ is smaller, which should again be related to the longer time scale per period and the larger contribution of Brownian motion which assists the random sliding of layers. This is concurrent to experimental rheology [8] and simulations [36] of hard sphere glasses where yield strain increases with frequency. At the smallest $Pe_\omega$ there is a minimum yield strain, being the least amount of distortion needed to allow a layer to escape using Brownian motion. Similarly for the highest $Pe_\omega$ there is a maximum yield strain which corresponds to the largest allowed distortion before the crystal layer is forced to slide to the next layer position.

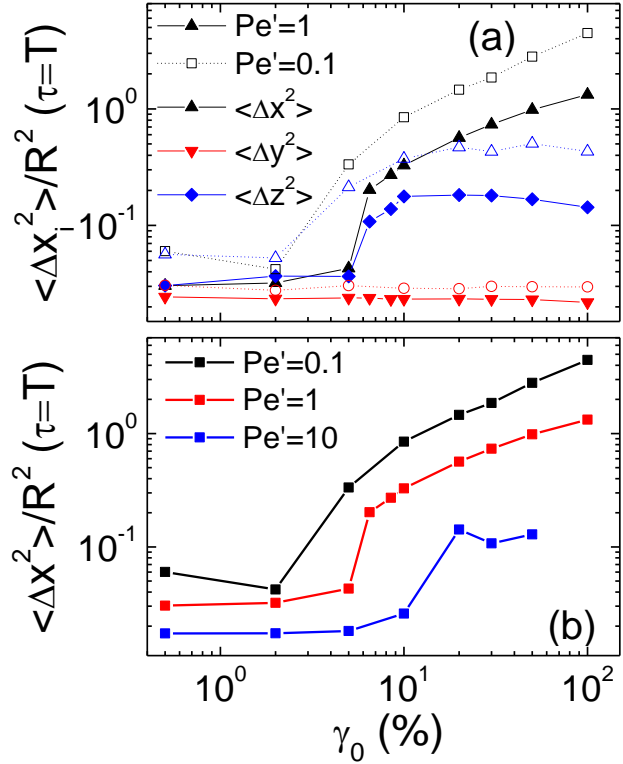

Figure 8: Mean squared displacements vs strain for a crystal parallel to shear at $\varphi=0.62$ for a time of a single oscillation period in a) all axis and two different $Pe_\omega$ (0.1 and 1) and b) only the x axis and three different $Pe_\omega$ (0.1, 1 and 10).

## Parallel vs. Perpendicular Crystals under Oscillatory Shear
### Parallel vs Perpendicular: Stresses

As discussed previously, in oscillatory shear experiments of concentrated hard spheres with non-rotational parallel plates, FCC crystals with the close packed direction parallel (fig. 1a) and perpendicular (fig. 1b) to shear have been found to occur at different strain regimes [39]. Starting from an amorphous configuration, low strain amplitudes produce a crystal with a distribution of orientation around a preferred orientation of the closed packed direction perpendicular to shear (fig. 1b). Higher strains of $\gamma_0 > 50\%$ reorient the existing crystal and produce a macroscopic crystal with the closed packed direction parallel to shear (fig. 1a). By lowering the strain with this specific structure, the system does not revert back to being amorphous or to the crystal perpendicular to shear. Previous work of [42] has discussed that this orientation change occurs through energy changes due to geometrical constraints. This work explains these orientation changes in terms of stresses and displacements under shear and comments on the lack thereof in the experimental observations on rotational geometries.

In figure 9 we compare the oscillatory stress response in Lissajous curves, stress versus strain within the period, of the crystals with parallel and perpendicular to shear direction for low and high strains at $Pe_\omega=1$ for a $\varphi=0.62$. The shapes of the curves are complicated, reflecting both the monocrystallinity and the yield properties of the system. At the low strain regime (fig. 9a) the two crystal orientations exhibit similar Lissajous shapes, with a linear elastic stress slope at the maxima of strain being almost equal. Beyond the simple elastic response, at the higher rates of the oscillation, the crystals reveal smaller stresses for the crystal perpendicular to shear. At the high strain regime (fig. 9b), the elastic response at the strain maxima is again similar, however, the higher rates of the oscillation show very different behaviors. The crystal parallel to shear shows stresses with decaying oscillations around a rate independent stress until strain reversal. In contrast, the crystal perpendicular to shear produces a strong increase of stress, much higher than the crystal parallel to shear, which then shows an oscillating decay to a large, rate dependent stress value.

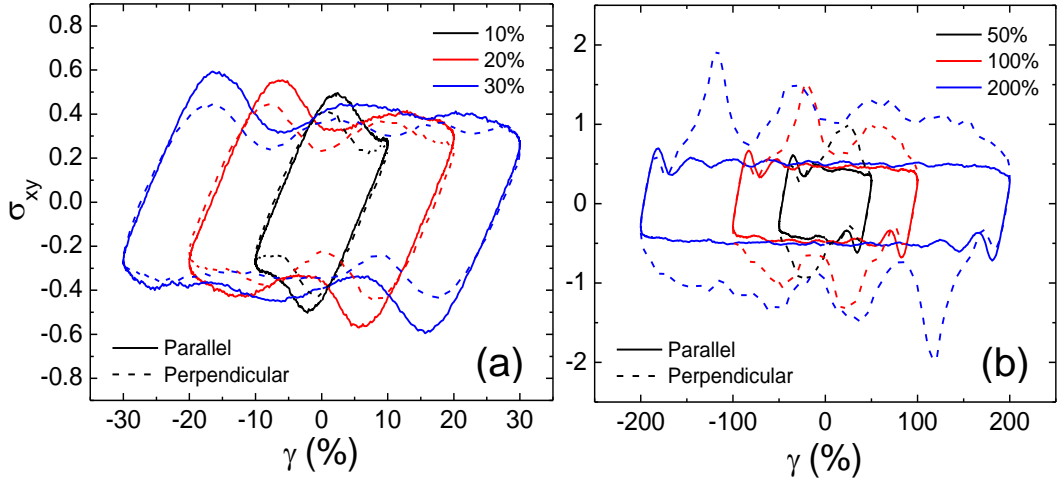

Figure 9: Lissajous (Stress-Strain) curves of crystals parallel and perpendicular to shear for $\varphi=0.62$ at $Pe_\omega=1$ for a) low nonlinear strains of 10%, 20% and 30% and b) high nonlinear strains of 50%, 100% and 200%.

Therefore, at the start of the nonlinear regime (low strains, figure 9a), the crystal parallel to shear exhibits larger stresses than the crystal perpendicular to shear, whereas for larger strains not only is this reversed but the crystal perpendicular to shear produces very strong stress peaks (figure 9b). This is correlated to the perpendicular to parallel orientation change observed in experiments. It seems that the smaller stresses for small strains allow a preferential perpendicular orientation to develop, albeit with a certain distribution, possibly because the difference is small. However, at larger strains this orientation gives prohibitive stresses, causing a reorientation to the crystal parallel to shear. If strains are then reduced, although the perpendicular crystal flows slightly easier, there is no driving force to cause the system to change orientations. Additionally, the linear regime stresses between the two orientations, although not shown, are identical as expected from simple FCC elasticity tensors [67]. Thus, the elastic slopes at the maxima of strain seen for both orientations in fig. 9 are equal and represent within those boundaries a linear elastic deformation of the crystal structures. It is important to note that the linear elastic stresses for such crystals are not generally equal in all directions.

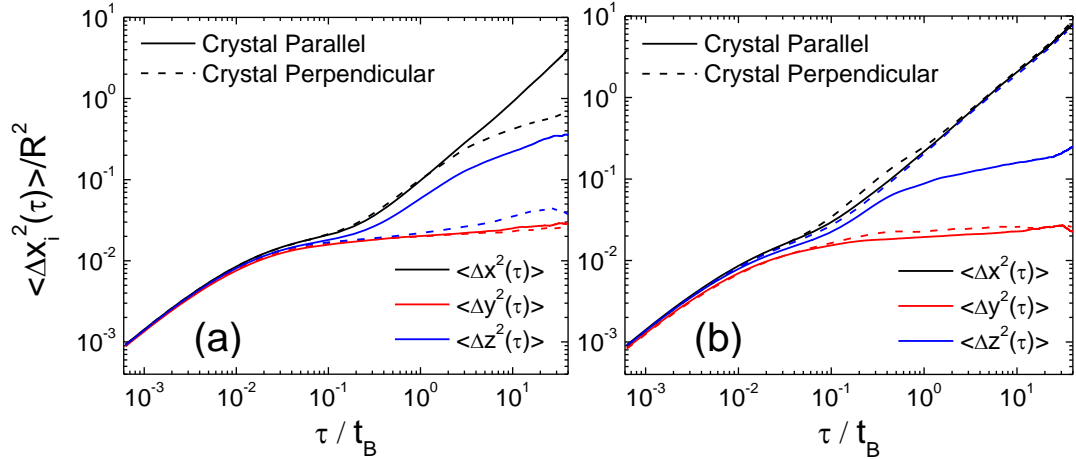

Figure 10: Mean squared particle displacements from crystals parallel and perpendicular to shear at $\varphi=0.62$ under oscillatory shear with a $Pe_\omega=1$ at a) $\gamma_0 = 20\%$ and b) $\gamma_0 = 100\%$.

**Parallel vs Perpendicular: Mean Squared Displacements**

Although the stresses make a clear point for the driving force behind the change of orientation, there may be more information hidden in the displacements. The time dependence of the MSDs under shear for the two orientations for $Pe_\omega=1$ at low $\gamma_0=20\%$ and high $\gamma_0=100\%$ can be seen in fig. 10. While the in-cage/short time displacements of the two crystals are equal between crystals, for low and high strains, at longer times they show differences. Similarly to the stress findings of the fig. 9, the longtime MSDs (t>T) for the crystal perpendicular to shear are in average less at low strains, while found to be larger at high strains. More specifically compared to its parallel counterpart, at low strains $\langle\Delta x^2\rangle$ and $\langle\Delta z^2\rangle$ become less for the perpendicular crystal and for high strains they increase. In both cases $\langle\Delta y^2\rangle$ remains the same in both orientations. Interestingly, $\langle\Delta x^2\rangle$ at $\gamma_0=20\%$ does not become diffusive at long time scales.

The MSD strain dependence of the orientations at $Pe_\omega=1$ is shown in fig. 11a along with the MSD calculated by subtracting neighbor motion in 11b. Starting from small $\gamma_0$, the crystal with perpendicular orientation shows a high $\langle\Delta x^2\rangle$ at smaller strains than the parallel one, thus exhibiting a smaller yield strain. Until $\gamma_0=30\%$, the displacements in the perpendicular crystal are smaller with $\langle\Delta z^2\rangle$ being constant, while for $\gamma_0>50\%$, they sharply increase above those of the parallel crystal. The MSD relative to the neighbors (fig. 11b) are similar for both orientations up to $\gamma_0=50\%$, but reveal an abrupt increase for the perpendicular crystal above this strain.

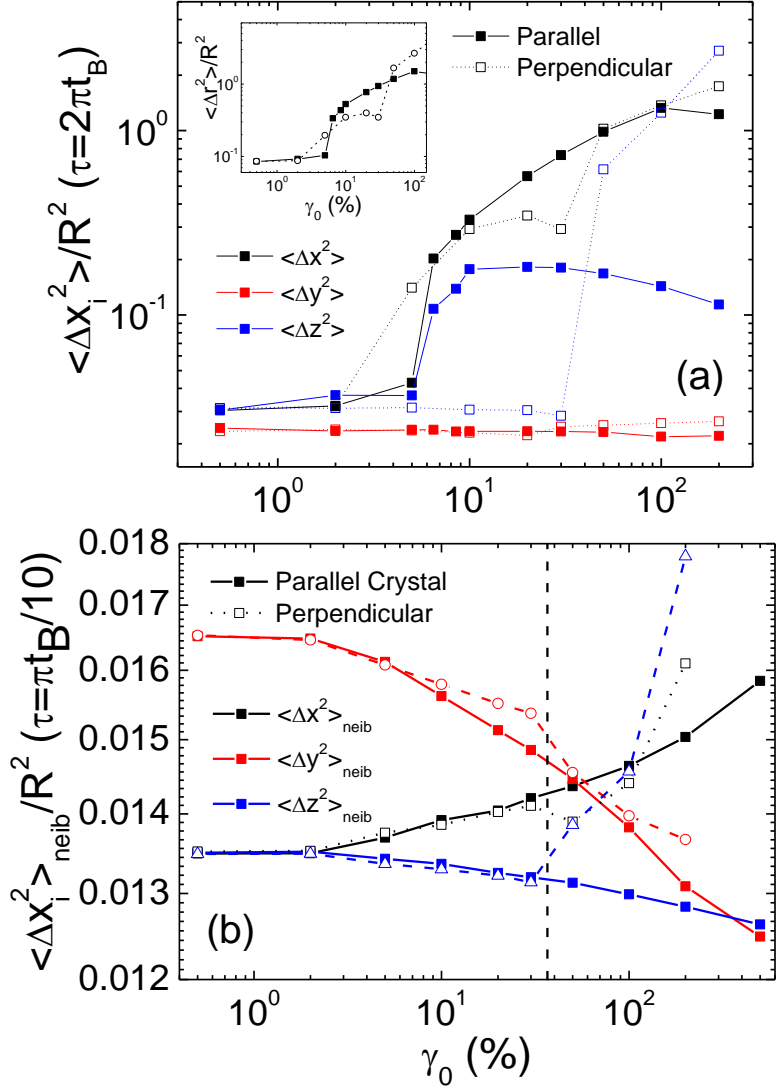

Figure 11: Mean squared displacements versus applied strain of crystals parallel and perpendicular to shear at $\varphi=0.62$ and $Pe_\omega=1$ a) for a time slice of $2\pi t_B$ (1 period), inset shows $\langle\Delta r^2(T)\rangle$ versus $\gamma_0$, b) the displacements relative to the same layer cage at a time of $\pi t_B/10$ (where in-cage motions dominate), while the vertical dashed line in (b) denotes the abrupt change in displacements of the perpendicular crystal.

The displacements of the different crystal orientations under shear follow the findings of the stresses from fig. 9. The displacements of the perpendicular crystal in small strains ($\gamma_0<50\%$) are less than the parallel, while the stresses are also less. When strains are increased above a certain point ($\gamma_0\geq50\%$) the crystal perpendicular to shear exhibits much higher displacements and stresses, additionally giving rise to crystal instabilities with the motion relative to neighbors being severely increased. This is an important indicator of instability as internal stresses are high enough to disrupt the internal layer arrangement.

**Parallel vs Perpendicular: Dynamics and Stresses within the Period**

Figures 12 and 13 show the stresses and instantaneous displacements within the period for the crystals parallel and perpendicular to shear for a low strain of 20% (fig. 12) and a high strain of 100% (fig. 13). As discussed earlier for low strains (fig. 9a), the two orientations have similar stress features although the perpendicular crystal exhibits lower absolute values. The instantaneous displacements at low strains also show similar features in the x and y axis as shown in 7a. However there is a marked difference in the z axis, where there are strong minima for the perpendicular crystal. For the high strains of fig. 13, the stresses are shown as in fig. 9b, where the crystal perpendicular to shear shows abrupt stress peaks. The high strain displacements of the crystal parallel to shear mimic the low strain behavior, although now the z axis shows significant changes. The perpendicular crystal in this case shows that the displacements are only anomalous in the z axis (vorticity).

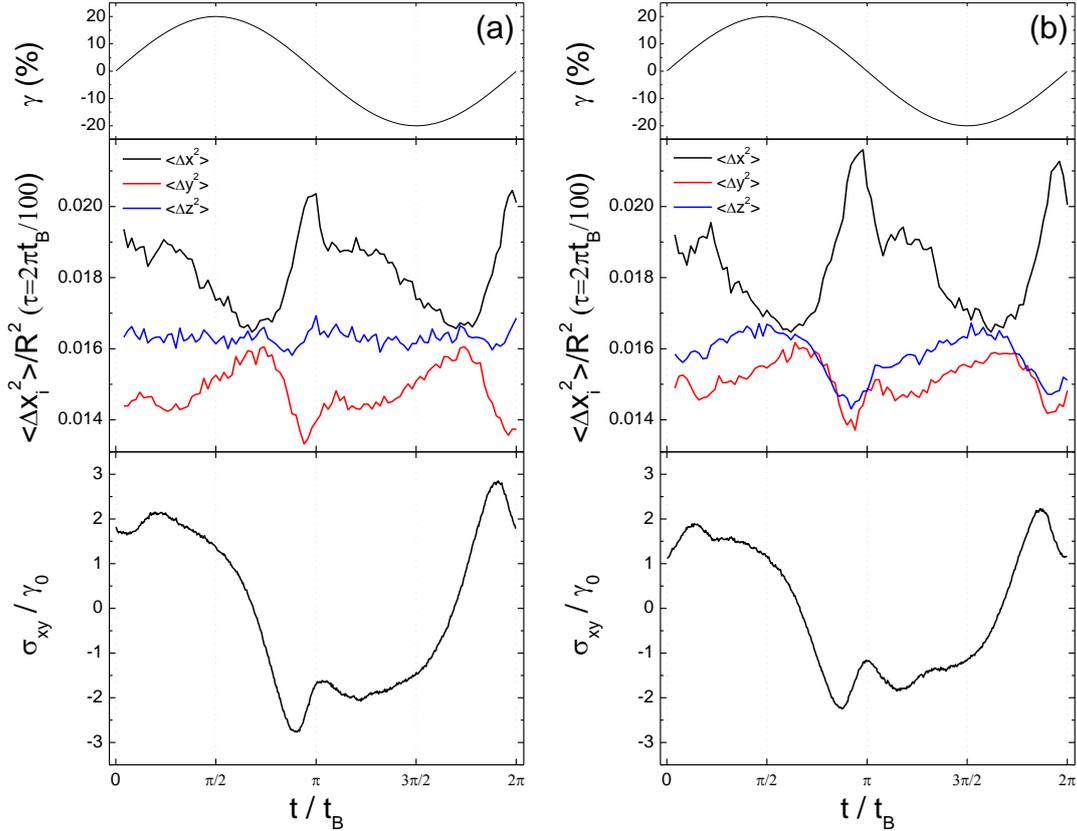

Figure 12: Mean squared displacements at $\tau=2\pi t_B/100$ ("instantaneous" MSD) for each axis at $\varphi=0.62$, $Pe_\omega=1$ and $\gamma_0=20\%$ within the period (top) and normalized stress (bottom) versus time of a) the crystal parallel to shear and b) the crystal perpendicular to shear.

Going back to figure 1, we schematically show the sliding behavior under shear of the parallel (fig.1a) and perpendicular (fig.1b) crystals. The layers of the crystal parallel to shear have only one slipping angle relative to the shear direction, $\pi/6$. However, the layers of the crystal perpendicular to shear have two slipping angles, the first one having a zero value and the second one being $\pi/3$. While even a zero angle would need a finite amount of stress to allow sliding of layers, as the angle is increased, the stress needed to slide the layers is also increased. Thus at low nonlinear strains the crystal perpendicular to shear has lower stresses (zero angle) and smaller displacements.

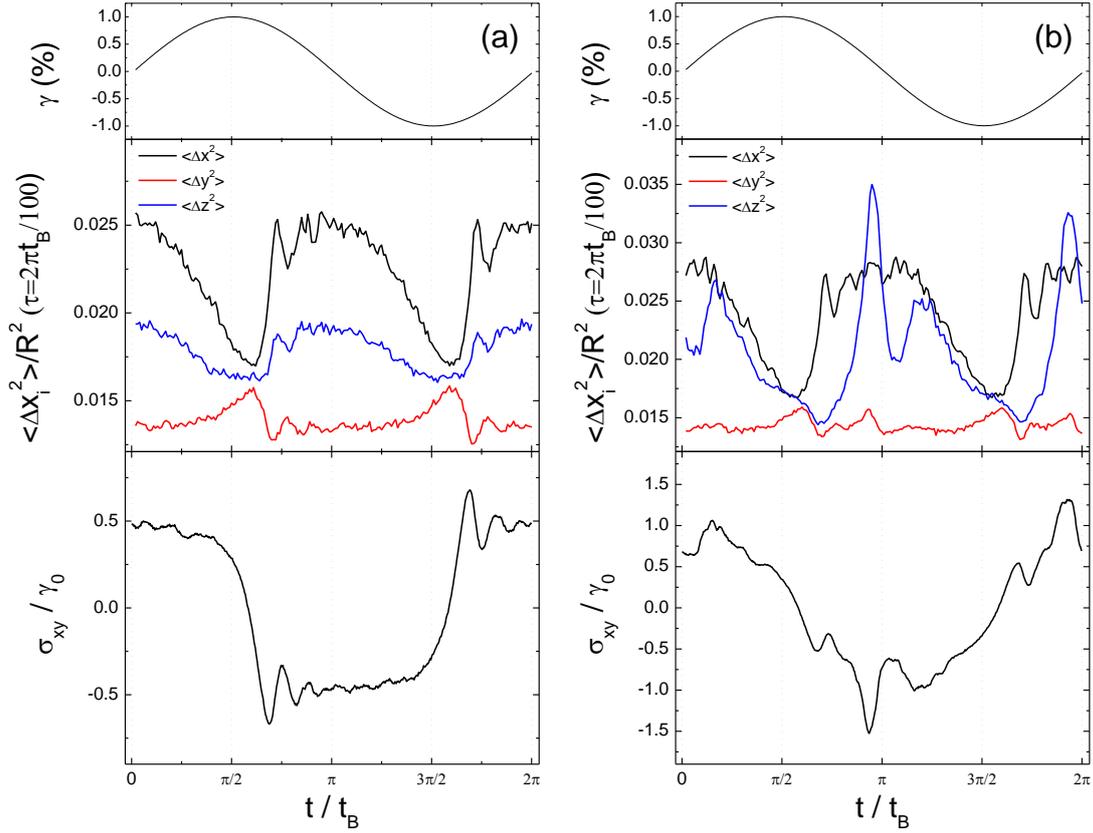

Figure 13: Mean squared displacements at $\tau=2\pi t_B/100$ ("instantaneous" MSD) for each axis at $\varphi=0.62$, $Pe_\omega=1$ and $\gamma_0=100\%$ within the period (top) and normalized stress (bottom) versus time of a) the crystal parallel to shear and b) the crystal perpendicular to shear.

As strain is increased, the perpendicular crystal is forced to explore an area of much larger stress ($\pi/3$ angle), which causes displacement instabilities in the z axis (figs 10, 11, 12 and 13). Experimentally these instabilities force the crystal to reorient parallel to shear. In that case a smaller sliding angle ($\pi/6$) is involved, which allows smoother layer motion with less stress. The crystal experimentally remains in this orientation for all strains, as there are no instabilities. In real systems however, at very high strains the crystal is finally destroyed by forces not taken into consideration as will be discussed. The smaller yield strain of the perpendicular crystal is also likely due to the initial zero angle slipping mechanism. An interesting observation can be made on the sub-diffusive motion seen for the perpendicular crystal at $\gamma_0=20\%$ (fig. 10a), showing confinement in the layer motion. As may be gleaned from the scheme of fig. 1a, the crystal will have difficulty slipping past the initial zero angle, leading to longtime confinement of the crystal layers and forcing collective motion of the crystal layers.

From the simulations we have found that the crystal parallel to shear is stable at all strains which is congruent with experimental observations [41]. The crystal perpendicular to shear is stable only at small non-linear strains, while showing high instabilities (in the y axis) at higher strains. It was also found that the crystal perpendicular to shear exhibits less stresses and MSDs than the crystal parallel to shear at low strains. Thus we conclude that the shear induced crystal experimentally is created first with a perpendicular orientation (less displacements and stresses), which then changes to parallel (reorientation due to instabilities in the perpendicular one). Since the crystal parallel to shear is still very efficient in reducing stresses under shear, crystal does not revert back to the perpendicular orientation at low strains.

Even at the highest strains (500%) the simulated crystal parallel to shear shows no instabilities (see SI). The slipping layer is such that whatever strain or $Pe_\omega$ is applied, the crystal is stable. At higher rates and strains there are actually less instabilities, as there is less Brownian motion to complicate the shear alignment of the crystal. The experimental instabilities for the parallel crystal, causing destruction of the crystal at high strains, are possibly hydrodynamic in origin [54], or as recently put forth, frictional in nature[68]. The lubrication or frictional stresses, which are not implemented in the Brownian dynamics code, could cause jamming of particles and subsequent breaking of the crystals. However, the instabilities of the crystal perpendicular to shear are spatial in origin (excluded volume) and are thus captured by the Brownian Dynamics simulations.

The lack of any perpendicular crystal in the rheometer cone-plate geometry [23] is possibly due to the curvature in the x-z plane, as there may be higher energies/stress required for creation of crystal. Additionally, the reason there is no full crystallization at lower strains, however long the elapsed time of shear, may be related to the experimental finding that in a parallel plate shear cell there is a distribution of crystal orientations around the perpendicular at low strains [39]. Since the perpendicular crystal is unable to form in the rotational geometry, only a small portion of parallel crystal is allowed, which by increasing the strain grows into a full crystal.

## Conclusions/Summary

With the use of Brownian Dynamics simulations, we have examined concentrated hard sphere suspensions under oscillatory shear with the purpose of understanding the route to shear-induced crystallization. Through analyses of particle motions and stresses averaged over the period, but also examined within the period, our work shows that the processes of shear induced ordering in concentrated suspensions not only minimizes the energy of the system by bringing it to an ordered state, but also leads to minimization of particle stresses as well as particle motions.

### Glass vs. crystal at rest

For the amorphous and crystalline structures at rest we found that although the short time MSD for the crystal is higher than that of the glass, the dynamic heterogeneities in the glass may allow increase of the MSD above those of the crystal for longer times. Thus, while in the crystal particles are completely confined in their cage, with the MSD showing a clear plateau, in the glass large scale out of cage motion is observed via dynamic heterogeneities. When approaching the limit of random close packing (≈0.66) the crystal MSD should always be larger that of a same φ glass because of the larger distance from maximum packing (≈0.74).

**Glass vs. parallel crystal under oscillatory shear**

When comparing the displacements of the crystal oriented parallel to shear and the glass under oscillatory shear we concluded that, due to the sliding layers the crystal has anisotropic displacements compared to the isotropic displacements found in the glass. The crystal displacements are due to cooperative motion of x-z layers of particles sliding over each other, although high rates cause cage elongation in the shear direction seen as an increase of $<\Delta x^2>$ relative to the x-z neighbors at long times. The yield strain of the crystal is found to be less than the glass, due to sliding layer motion. Past the yield point, the longtime shear induced displacements of the glass are found to be larger, while stresses are highly correlated to instantaneous shear induced displacements in both crystal and glass. We conclude that during the experimental shear of a monodisperse glass, the large out of cage displacements allow the system to explore the energy landscape and find the minima in energy, stresses and displacements by configuring particles into a crystal oriented parallel to shear, simultaneously reducing non-Gaussian behavior in the particle motions.

**Parallel crystal under oscillatory shear**

Through the examination of the parallel crystal under oscillatory shear, we have found that the MSD of the crystal increases with decreasing $Pe_\omega$, although showing qualitatively similar displacements as a function of applied strain. The yield strain of the crystal is less for lower oscillation frequencies as Brownian motion has time to act upon the system under shear and allow the sliding of the crystal layers.

**Parallel vs. perpendicular crystals under oscillatory shear**

The experimental transition from glass to crystal with increasing strain as well as the selection of crystal orientation is governed by the minimization of stresses and displacements in the hard sphere system. Similarly, the low strain perpendicular and high strain parallel to shear crystals may occur due to minimization of stresses and displacements. The crystal perpendicular to shear exhibits low stresses and MSD at low nonlinear strains (10%-20%), while the one parallel to shear shows slightly larger stresses in the same regions. At higher strains, the sliding layers of the parallel crystal are more efficient in minimizing stresses and the perpendicular crystal is no longer stable. Thus starting from an amorphous glassy system, at low strains a crystal perpendicular to shear is formed, which becomes unstable at larger strains and reorients into a crystal parallel to shear.

# Acknowledgments

N.K. has been supported by the Greek General Secretariat for Research and Technology (Basic Research Program PENED,-03ED566) and Horizon 2020 funding, through H2020-MSCA-IF-2014, ActiDoC No. 654688, from the European Union (EU).